\documentclass[journal,twocolumn]{IEEEtran}
\usepackage{amsfonts}
\usepackage{amsmath}
\usepackage{amssymb}
\usepackage{array}
\usepackage{bm}
\usepackage{booktabs}
\usepackage{cite}
\usepackage{color}
\usepackage{dsfont}
\usepackage{enumerate}
\usepackage{epsfig}
\usepackage{epstopdf}
\usepackage{algorithm}
\usepackage{algorithmic}
\usepackage{caption}

\usepackage{tabu}

\newtheorem{proposition}{Proposition}

\def\BD{\begin{displaymath}}
\def\BE{\begin{equation}}
\def\BEA{\begin{eqnarray}}
\def\BEAs{\begin{eqnarray*}}
\def\ED{\end{displaymath}}
\def\EE{\end{equation}}
\def\EEA{\end{eqnarray}}
\def\EEAs{\end{eqnarray*}}

\def\bA{{\bf A}}

\def\bB{{\bf B}}

\def\bC{{\bf C}}

\def\bD{{\bf D}}
\def\bd{{\bf d}}
\def\bE{{\bf E}}
\def\be{{\bf e}}
\def\bF{{\bf F}}
\def\bG{{\bf G}}

\def\bH{{\bf H}}

\def\bI{{\bf I}}

\def\bL{{\bf L}}

\def\bn{{\bf n}}
\def\bP{{\bf P}}
\def\bp{{\bf p}}
\def\bQ{{\bf Q}}

\def\bs{{\bf s}}

\def\bU{{\bf U}}
\def\bu{{\bf u}}

\def\bw{{\bf w}}

\def\by{{\bf y}}

\def\bz{{\bf z}}

\def\b_eta{\mbox{\boldmath $\eta$}}

\def\bnu{\mbox{\boldmath $\nu$}}

\def\diag { {\rm diag} }

\def\trace{ {\rm tr} }

\def\L2{L^{2}[-{\pi \over 2} , {\pi \over 2}]}

\def\bnu{\bnu}

\graphicspath{{./}{Figures/}}

\begin{document}
%
\title{Linear Precoding for Fading Cognitive Multiple Access Wiretap Channel with Finite-Alphabet Inputs}
\author{Juening~Jin, ~Chengshan~Xiao,~\IEEEmembership{Fellow,~IEEE,} ~Meixia~Tao,~\IEEEmembership{Senior Member,~IEEE,}
\\~Wen~Chen,~\IEEEmembership{Senior Member,~IEEE}
\thanks{The work of C. Xiao was supported in part by the U.S. National Science Foundation under Grants ECCS-1231848 and ECCS-1539316. The work of M. Tao was supported by the National Natural Science Foundation of China under Grant 61322102. The work of W. Chen was supported in part by the national 973 project under Grant 2012CB316106 and the national 863 project under Grant 2015AA01A710. This work has been carried out while Mr. Juening Jin is visiting Missouri University of Science and Technology. Part of the material in this paper was presented at the IEEE International Conference on Communications, Kuala Lumpur, Malaysia, 2016. }
\thanks{J. Jin, and M. Tao are with the Department of Electronic Engineering,
Shanghai Jiao Tong University, Shanghai 200240, China (E-mail:
jueningjin@gmail.com; mxtao@sjtu.edu.cn).}
\thanks{C. Xiao is with the Department of Electrical and Computer Engineering,
 Missouri University of Science and Technology, Rolla, MO  65409, USA
 (E-mail: xiaoc@mst.edu).}
\thanks{W. Chen is with the Shanghai Key Laboratory of Navigation and Location
Based Services, Shanghai Jiao Tong University, Shanghai 200240, China,
and also with the School of Electronics Engineering and Automation, Guilin
University of Electronics Technology, Guilin 541004, China (E-mail: wenchen@sjtu.edu.cn).}
 \thanks{Copyright \copyright\; 2015 IEEE. Personal use of this material is permitted. However, permission to use this material for any other purposes must be obtained from the IEEE by sending a request to pubs-permissions@ieee.org.}
 }


\maketitle

\vspace{-1.7cm}
\begin{abstract}
We investigate the fading cognitive multiple access wiretap channel (CMAC-WT), in which two secondary-user transmitters (STs) send secure messages to a secondary-user receiver (SR) in the presence of an eavesdropper (ED) and subject to interference threshold constraints at multiple primary-user receivers (PRs). We design linear precoders to maximize the average secrecy sum rate for multiple-input multiple-output (MIMO) fading CMAC-WT under finite-alphabet inputs and statistical channel state information (CSI) at STs. For this non-deterministic polynomial time (NP)-hard problem, we utilize an accurate approximation of the average secrecy sum rate to reduce the computational complexity, and then present a two-layer algorithm by embedding the convex-concave procedure into an outer approximation framework. The idea behind this algorithm is to reformulate the approximated average secrecy sum rate as a difference of convex functions, and then generate a sequence of simpler relaxed sets to approach the non-convex feasible set. Subsequently, we maximize the approximated average secrecy sum rate over the sequence of relaxed sets by using the convex-concave procedure. Numerical results indicate that our proposed precoding algorithm is superior to the conventional Gaussian precoding method in the medium and high signal-to-noise ratio (SNR) regimes.
\end{abstract}


%

\section{Introduction}
Spectrum sharing has been widely recognized as a promising technology to improve the utilization efficiency of the limited spectrum resources in cognitive radio networks \cite{goldsmith2009breaking}. In a spectrum sharing cognitive radio network, unlicensed secondary users are allowed to communicate concurrently with licensed primary users over the same bandwidth as long as the interference power at primary-user receivers is kept below a given threshold. Related works in \cite{zhang2008joint,zhang2009weighted} considered the weighted sum rate optimization in cognitive radio networks with interference threshold constraints.

Meanwhile, due to the open and broadcast nature of radio propagation, such a spectrum sharing may cause security problems because all kinds of wireless equipments are able to overhear the licensed spectrum. Therefore, security is a critical issue in cognitive radio networks. Traditionally, security of a network has been entrusted in the network layer through cryptography and authentication, which often require additional system complexity for key generation and complex encryption/decryption algorithms \cite{chan2003random}.

In recent years, there has been growing interest in physical-layer security that enables secure communication over the physical layer. Physical-layer security or information-theoretic security originated from Shannon's notion of perfect secrecy \cite{shannon1949communication}. It was first studied in wiretap channel by Wyner \cite{wyner1975wire} and later in broadcast channel with confidential messages by Csisz\'ar and K\"{o}rner \cite{csiszar1978broadcast}. The study of physical-layer security is then extended to several multiuser communication scenarios. In \cite{tekin2008gaussian}, the authors introduced the degraded Gaussian multiple access wiretap channel, where an additional eavesdropper is able to access to the multiple access channel output via a degraded wiretap channel. In \cite{tekin2008general}, an achievable secrecy rate region with Gaussian inputs was proposed for the non-degraded Gaussian multiple access wiretap channel, and the power allocations maximizing the corresponding secrecy sum rate were also determined. Related works in \cite{geraci2012secrecy,geraci2013large,yang2014confidential,yang2014joint,hanif2014linear} further investigated linear precoding designs that maximize the secrecy (sum) rate in other multiple-input multiple-output (MIMO) multiuser channels.

The precoding designs in \cite{geraci2012secrecy,geraci2013large,yang2014confidential,yang2014joint,hanif2014linear} require instantaneous channel state information (CSI) of both legitimate receivers and eavesdroppers. However, such a requirement is over-optimistic for fast fading channels, of which the channel coherence time may be shorter than the feedback delay caused by channel estimation. In this case, when the instantaneous CSI is arrived at transmitters, the channel state has already changed. Therefore, it is more realistic to exploit channel statistics at transmitters for precoding design, due to its much slower changes compared with instantaneous CSI.

Furthermore, the results in\cite{geraci2012secrecy,geraci2013large,yang2014confidential,yang2014joint,hanif2014linear} rely on the ideal assumption of Gaussian inputs. Although Gaussian inputs are proven to be capacity achieving in a variety of Gaussian channels, they are hardly implemented in practice. It is well known that practical inputs are drawn from finite constellation sets such as phase-shift keying (PSK), pulse-amplitude modulation (PAM), or quadrature amplitude modulation (QAM). More importantly, the common approach that designs linear precoder in a MIMO system under Gaussian inputs and then apply it to the practical system may lead to significant performance loss \cite{lozano2006optimum,xiao2008mutual}. Therefore, the precoding design with finite-alphabet inputs has drawn increasing research interest in recent years \cite{perez2010mimo, xiao2011globally,wang2011linear,zeng2012globally,zeng2012linear,bashar2012secrecy,wu2012linear,harshan2013novel,vishwakarma2013decode,vishwakarma2014power,girnyk2014large,zeng2015Cwiretap}.

As illustrated in Fig. 1, we consider the underlay cognitive multiple access wiretap channel (CMAC-WT), where two secondary-user transmitters (STs) communicate with one secondary-user receiver (SR) in the presence of an eavesdropper (ED) and subject to interference threshold constraints at primary-user receivers (PRs). Each node in the system is equipped with multiple antennas. To the best of our knowledge, this is a general model that has not been addressed yet. We design linear precoding matrices to achieve the maximum average secrecy sum rate under finite-alphabet inputs and statistical CSI at STs. The problem setting is much closer to practical systems because it targets finite-alphabet inputs directly and exploits statistical CSI of fading channels. However, this problem is extremely difficult to solve due to two reasons: First, the computational complexity for evaluating the average secrecy sum rate is prohibitively high. Second, and more importantly, the optimization problem itself is a non-convex and non-deterministic polynomial time (NP)-hard problem.

A subset of non-convex optimization, which is called the difference of convex functions (DC) optimization, has been studied extensively by exploiting its underlying structure\cite{horst1996global,tao1988duality,yuille2003concave}. DC optimization aims to maximize a DC function under some DC constraints. In \cite{horst1996global}, a basic outer approximation framework was proposed for solving DC problems. In \cite{tao1988duality}, a new DC algorithm was introduced by exploiting the duality theory of DC optimization. In \cite{yuille2003concave}, the authors presented the convex-concave procedure, which can be regarded as a special case of the algorithm in \cite{tao1988duality}. Since any twice continuously differentiable function is a DC function \cite{horst1996global}, our linear precoding problem is a DC optimization problem. However, no practical algorithm is known to construct a DC decomposition for arbitrary twice continuously differentiable function. Moreover, if we do not carefully design the DC representation of the average secrecy sum rate, the algorithms in \cite{horst1996global,tao1988duality,yuille2003concave} suffer from very slow convergence\cite{ferrer2009improving}. Therefore, the DC representation is a main factor that affect the performance of DC algorithms.

We solve our problem efficiently by combining the convex-concave procedure with an outer approximation framework. We first exploit an accurate approximation of the average secrecy sum rate to reduce the complexity, and then reformulate the approximated average secrecy sum rate as a DC function. Subsequently, we generate a sequence of relaxed sets, which can be expressed explicitly as the union of convex sets, to approach the non-convex feasible set. In this way, near optimal precoders are obtained by maximizing the approximated average secrecy sum rate over these convex sets.
\begin{figure}[!ht]
  \begin{center}
  \includegraphics[scale=0.3]{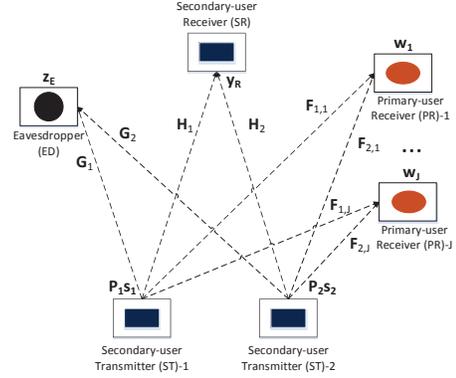}\\
  \caption{System model of the fading cognitive multiple access wiretap channel.}
  \end{center}
  \vspace{-0.6cm}
\end{figure}
Numerical results show that when considering finite-alphabet inputs, our proposed algorithm significantly outperforms the conventional Gaussian precoding method, which designs precoding matrices to maximize the average secrecy sum rate under Gaussian inputs, in the medium and high signal-to-noise ratio (SNR) regimes.

The rest of this paper is organized as follows. Section II introduces the system model and formulates the linear precoding problem, Section III develops a numerical algorithm to maximize the average secrecy sum rate under finite-alphabet inputs and statistical CSI, Section IV presents several numerical results and Section V draws the conclusion.

\emph{Notations}: Boldface lowercase letters, boldface uppercase letters, and calligraphic letters are used to denote vectors, matrices and sets, respectively. The superscripts $(\cdot)^T$ and $(\cdot)^H$ represent transpose and Hermitian operations, respectively. $[\cdot]^+$ denotes $\max(\cdot,0)$; $\diag(\cdot)$ represents a block diagonal matrix whose diagonal elements are matrices. $\mathrm{tr}(\cdot)$ is the trace
of a matrix; $\mathrm{vec}(\cdot)$ is a column vector formed by stacking the columns of a matrix; $\|\cdot\|$ denotes the Euclidean norm
of a vector; $\bA\otimes\bB$ is the Kronecker product of two matrices $\bA$ and $\bB$; $E(\cdot)$ represents the statistical
expectation; $\Re(\cdot)$ and $\Im(\cdot)$ denote the real and image parts of a complex vector or matrix; $\geq$ and $\leq$ are defined component-wise. $\bI$ and $\bm{0}$ denote an identity matrix and a zero matrix, respectively, with appropriate dimensions; $\bA\succeq \bm{0}$ denotes the positive semidefiniteness of $\bA$. The symbol $\mathcal{I}(\cdot)$ represents the mutual information; $\log(\cdot)$ and $\ln(\cdot)$ are used for the base two logarithm and natural logarithm, respectively.

\section{System Model and Problem Formulation}
We consider the fading CMAC-WT depicted in Fig. 1. The $i$-th ST has $N_{\scriptscriptstyle T_{i}}$ antennas, $i=1,2$, the SR has $N_{\!\scriptscriptstyle R}$ antennas, the ED has $N_{\!\scriptscriptstyle E}$ antennas, and the $j$-th PR has $N_j$ antennas, $j=1,2,...,J$. The channel output at the SR, the ED and the $j$-th PR are, respectively, given by
\begin{align}
 \by_{\!\scriptscriptstyle R}&=\bH_{1}\bP_{\!1}\bs_{1}+\bH_{2}\bP_{\!2}\bs_{2}+\bn_{\scriptscriptstyle R}\nonumber \\
 \bz_{\scriptscriptstyle E}&=\bG_{1}\bP_{\!1}\bs_{1}+\bG_{2}\bP_{\!2}\bs_{2}+\bn_{\scriptscriptstyle E} \nonumber \\
 \bw_{\!j}&=\bF_{\!1,j}\bP_{\!1}\bs_{1}+\bF_{\!2,j}\bP_{\!2}\bs_{2}+\bn_j,\; j=1,2,...,J
\end{align}
where $\bH_{i}$, $\bG_{i}$ and $\bF_{\!i,j}$ are complex channel matrices from the $i$-th ST to the SR, the ED, and the $j$-th PR, respectively; $\bP_{\!i}$ is the linear precoding matrix at the $i$-th ST, $i=1,2$; $\bs_{i}$ is the input data vector at the $i$-th ST with zero-mean and covariance $E_{\bs_{i}}[\bs_{i}\bs_{i}^{H}]\!=\!\bI$, $i=1,2$;  $\bn_{\scriptscriptstyle R}$, $\bn_{\scriptscriptstyle E}$ and $\bn_j$ are independent and identically distributed (i.i.d.) zero-mean circularly symmetric complex Gaussian noises with covariance matrix $\sigma_{\!\scriptscriptstyle R}^2\bI$, $\sigma_{\!\scriptscriptstyle E}^2\bI$ and $\sigma_{\!j}^2\bI$, respectively.

The channel matrices considered in this paper are modeled as \cite{xiao2004discrete}
\begin{align}
&\bH_{i}=\bm{\Phi}^{\frac{1}{2}}_{\!h}\tilde{\bH}_{i}\bm{\Psi}^{\frac{1}{2}}_{\!h_{i}}, \quad i=1,2\nonumber\\
&\bG_{i}=\bm{\Phi}^{\frac{1}{2}}_{\!g}\tilde{\bG}_{i}\bm{\Psi}^{\frac{1}{2}}_{\!g_{i}}, \quad i=1,2\nonumber\\
&\bF_{\!i,j}=\bm{\Phi}^{\frac{1}{2}}_{\!f_{j}}\tilde{\bF}_{\!i,j}\bm{\Psi}^{\frac{1}{2}}_{\!f_{i,j}}, \quad \forall{(i,j)}
\end{align}
where $\tilde{\bH}_{i}$, $\tilde{\bG}_{i}$ and $\tilde{\bF}_{\!i,j}$ are random matrices with i.i.d. zero-mean unit variance complex Gaussian entries; $\bm{\Phi}_{\!h}$, $\bm{\Phi}_{\!g}$ and $\bm{\Phi}_{\!f_{j}}$ are positive semidefinite receive correlation matrices of $\bH_{i}$, $\bG_{i}$ and $\bF_{\!i,j}$, respectively; $\bm{\Psi}_{\!h_{i}}$, $\bm{\Psi}_{\!g_{i}}$ and $\bm{\Psi}_{\!f_{i,j}}$ are positive semidefinite transmit correlation matrices of $\bH_{i}$, $\bG_{i}$ and $\bF_{\!i,j}$, respectively.

We assume that the SR has instantaneous channel realizations of $\{\bH_1,\bH_2\}$, the ED has instantaneous channel realizations of $\{\bG_1,\bG_2\}$, and STs only know the transmit and receive correlation matrices of $\{\bH_1,\bH_2,\bG_1,\bG_2,\bF_{i,j}, \forall (i,j)\}$ as well as the distributions of $\{\tilde{\bH}_1,\tilde{\bH}_2,\tilde{\bG}_1,\tilde{\bG}_2,\tilde{\bF}_{i,j}, \forall (i,j)\}$. Under these assumptions, the following secrecy sum rate is achievable \cite{tekin2008general}:
\begin{align*}
\big[\mathcal{I}&(\bs_{1},\bs_{2};\by_{\!\scriptscriptstyle R}|\bH)-\mathcal{I}(\bs_{1},\bs_{2};\bz_{\scriptscriptstyle E}|\bG)\big]^{+}\nonumber\\
&\!=\!\big[E_{\scriptscriptstyle\bH}\mathcal{I}(\bs_{1},\bs_{2};\by_{\!\scriptscriptstyle R}|\bH\!=\!\bar{\bH})\!-\!E_{\scriptscriptstyle\bG}\mathcal{I}(\bs_{1},\bs_{2};\bz_{\scriptscriptstyle E}|\bG\!=\!\bar{\bG})\big]^{+}
\end{align*}
where $\bH=[\bH_1,\bH_2]$, $\bG=[\bG_1,\bG_2]$; $\bar{\bH}$ and $\bar{\bG}$ represent the instantaneous channel realizations of $\bH$ and $\bG$, respectively. For notational simplicity, we omit the given channel realization condition in mutual information expressions and then the average secrecy sum rate can be expressed as
\begin{align}
R_{\mathrm{avg}}(\bP_{\!1},\bP_{\!2})=\big[E_{\scriptscriptstyle\bH}\mathcal{I}(\bs_{1},\bs_{2};\by_{\!\scriptscriptstyle R})-E_{\scriptscriptstyle\bG}\mathcal{I}(\bs_{1},\bs_{2};\bz_{\scriptscriptstyle E})\big]^{+}.
\end{align}

We maximize $R_{\mathrm{avg}}(\bP_{\!1},\bP_{\!2})$ subject to power constraints at STs and interference threshold constraints at PRs. The average transmit power conforms to the power constraint $\beta_{i}$:
\begin{align}\label{constraint1}
&E_{\scriptscriptstyle\bs_{i}}\trace\big(\bP_{\!i}\bs_{i}\bs_{i}^{H}\bP_{\!i}^H\big)=\trace\big(\bP_{\!i}^H\bP_{\!i}\big)\leq\beta_{i}, \quad i=1,2
\end{align}
and the average interference power at the $j$-th PR is limited by $\gamma_{j}$:
\begin{align}\label{constraint2}
\nonumber
\sum_{i=1}^2 &E_{\scriptscriptstyle\bs_{i},\bF_{\!i,j}} \Big[\trace\big(\bF_{\!i,j}\bP_{\!i}\bs_{i}\bs_{i}^{H}\bP_{\!i}^H\bF_{\!i,j}^H\big)\Big]\\  \nonumber
&=\sum_{i=1}^2 E_{\scriptscriptstyle \tilde{\bF}_{\!i,j}}\Big[\trace\big(\bP_{\!i}^H(\bm{\Psi}^{\frac{1}{2}}_{\!f_{i,j}})^{H}\tilde{\bF}_{\!i,j}^{H}\bm{\Phi}_{\!f_{j}}\tilde{\bF}_{\!i,j}\bm{\Psi}^{\frac{1}{2}}_{\!f_{i,j}}\bP_{\!i}\big)\Big]\\
&=\trace(\bm{\Phi}_{\!f_{j}})\cdot\sum_{i=1}^2 \trace\big(\bP_{\!i}^H\bm{\Psi}_{\!f_{i,j}}\bP_{\!i}
\big)\leq \gamma_{j}, \; \forall j.
\end{align}
The second equality in \eqref{constraint2} holds because each element of $\tilde{\bF}_{\!i,j}$ is i.i.d. complex Gaussian variable with zero-mean and unit variance, and $\tilde{\bF}_{\!i,j}$ is independent to $\bs_{i}$. Then the average secrecy sum rate maximization problem is formulated as
\begin{equation}\label{OPT1}
\begin{aligned}
& \underset{\bP_{\!1},\bP_{\!2}}{\mathrm{maximize}}
& & \quad R_{\mathrm{avg}}(\bP_{\!1},\bP_{\!2})\\
& \mathrm{subject \;to}
& & \quad\;\eqref{constraint1} \;\mathrm{and}\; \eqref{constraint2}.
\end{aligned}
\end{equation}

\section{Linear Precoding under Finite-Alphabet Inputs}

In this section, we solve problem \eqref{OPT1} under finite-alphabet inputs. We assume that each symbol of the input data vector $\bs_{i}$ is taken independently from an equiprobable discrete constellation with cardinality $M_{i}$, $i=1,2$. The average constellation-constrained mutual informations $E_{\scriptscriptstyle \bH}\mathcal{I}(\bs_1,\bs_2;\by_{\!\scriptscriptstyle R})$ and $E_{\scriptscriptstyle \bG}\mathcal{I}(\bs_1,\bs_2;\bz_{\scriptscriptstyle E})$ can then be expressed respectively as \cite{wang2011linear}
\begin{align}
E_{\scriptscriptstyle \bH}\mathcal{I}(\bs;\by_{\!\scriptscriptstyle R})=&\log N-\frac{1}{N}\sum_{m=1}^{N} E_{\scriptscriptstyle\bH,\bn_{\!\scriptscriptstyle R}}\bigg\{\log\sum_{k=1}^{N}\nonumber\\&\exp\Big(\frac{-\| \bH\bP\be_{mk}+\bn_{\scriptscriptstyle R}\|^2+\|\bn_{\scriptscriptstyle R}\|^2}{\sigma_{\!\scriptscriptstyle R}^{2}}\Big)\bigg\}\label{EMI1}
\end{align}
\vspace{-0.6cm}
\begin{align}
E_{\scriptscriptstyle\bG}\mathcal{I}(\bs;\bz_{\!\scriptscriptstyle E})=&\log N-\frac{1}{N}\sum_{m=1}^{N} E_{\scriptscriptstyle\bG,\bn_{\!\scriptscriptstyle E}}\bigg\{\log\sum_{k=1}^{N}\nonumber\\&\exp\Big(\frac{-\|\bG\bP\be_{mk}+\bn_{\scriptscriptstyle E}\|^2+\|\bn_{\scriptscriptstyle E}\|^2}{\sigma_{\!\scriptscriptstyle E}^{2}}\Big)\bigg\}\label{EMI2}
\end{align}
where $\bs\!=\![\bs_{1}^{T},\bs_{2}^{T}]^{T}$; $N$ is a constant, equals to $M_{1}^{N_{T_{1}}}M_{2}^{N_{T_{2}}}$; $\bP\!=\!\diag(\bP_1,\bP_2)$; $\be_{mk}$ is the difference between $\bd_m$ and $\bd_k$, with $\bd_m$ and $\bd_k$ representing two possible distinct signal vectors from $\bs$.

Obviously, the evaluation and optimization of the above average mutual informations is a difficult task. In order to obtain $E_{\scriptscriptstyle\bH}\mathcal{I}(\bs;\by_{\!\scriptscriptstyle R})$ and $E_{\scriptscriptstyle\bG}\mathcal{I}(\bs;\bz_{\scriptscriptstyle E})$, we need to  calculate expectations over $\bH$ and $\bG$ as well as $\bn_{\scriptscriptstyle R}$ and $\bn_{\scriptscriptstyle E}$. Unfortunately, these expectations have no closed-form expressions. Although we can use Monte Carlo method to estimate these expectations, the computational complexity is prohibitively high especially when the dimensions of $\bH$ and $\bG$ are large.

This difficulty can be mitigated by employing accurate approximations of \eqref{EMI1} and \eqref{EMI2}. Based on \cite{zeng2012linear}, $E_{\scriptscriptstyle\bH}\mathcal{I}(\bs;\by_{\!\scriptscriptstyle R})$ and $E_{\scriptscriptstyle\bG}\mathcal{I}(\bs;\bz_{\scriptscriptstyle E})$ can be approximated respectively as
\vspace{-0.1cm}
\begin{align}
\mathcal{I}_{\!\scriptscriptstyle A}(\bs;\by_{\!\scriptscriptstyle R})=&\log N-\frac{1}{N}\sum_{m=1}^{N}\log\sum_{k=1}^{N} \nonumber\\
&\prod_{q}\Big(1+\frac{h_q}{2\sigma_{\!\scriptscriptstyle R}^2}\be_{mk}^H\bP^H\bm{\Psi}_{\!h}\bP\be_{mk}\Big)^{-1}\label{3A1}\\
\mathcal{I}_{\!\scriptscriptstyle A}(\bs;\bz_{\scriptscriptstyle E})=&\log N-\frac{1}{N}\sum_{m=1}^{N}\log\sum_{k=1}^{N} \nonumber\\ &\prod_{q}\Big(1+\frac{g_q}{2\sigma_{\!\scriptscriptstyle E}^2}\be_{mk}^H\bP^H\bm{\Psi}_{\!g}\bP\be_{mk}\Big)^{-1}\label{3A2}
\end{align}
where $\bm{\Psi}_{\!h}=\diag(\bm{\Psi}_{\!h_1},\bm{\Psi}_{\!h_2})$ and
$\bm{\Psi}_{\!g}=\diag(\bm{\Psi}_{\!g_1},\bm{\Psi}_{\!g_2})$; $h_q$ and $g_q$ represent the $q$-th eigenvalue of $\bm{\Phi}_{\!h}$ and $\bm{\Phi}_{\!g}$, respectively. Approximations \eqref{3A1} and \eqref{3A2} are very accurate for arbitrary correlation matrices and precoders, and the computational complexity of \eqref{3A1} and \eqref{3A2} is several orders of magnitude lower than that of the original average mutual informations \cite{zeng2012linear}.

By replacing $R_{\mathrm{avg}}(\bP_{\!1},\bP_{\!2})$ with $[\mathcal{I}_{\!\scriptscriptstyle A}(\bs;\by_{\!\scriptscriptstyle R})-\mathcal{I}_{\!\scriptscriptstyle A}(\bs;\bz_{\scriptscriptstyle E})]^+$, problem \eqref{OPT1} can be approximated as
\begin{equation}\label{opt2}
\begin{aligned}
& \underset{\bP_1,\bP_2}{\mathrm{maximize}}
& & \big[\mathcal{I}_{\!\scriptscriptstyle A}(\bs;\by_{\!\scriptscriptstyle R})-\mathcal{I}_{\!\scriptscriptstyle A}(\bs;\bz_{\scriptscriptstyle E})\big]^{+} \\
& \mathrm{subject \;to}
& & \qquad\eqref{constraint1}\; \mathrm{and} \;\eqref{constraint2}.
\end{aligned}
\end{equation}

\subsection{Precoder vectorization}
We reformulate problem \eqref{opt2} into a vectorized form by employing the precoder vectorization technique \cite{xu2011joint,zeng2015Cwiretap}. This reformulation can better exploit the inherent structure of \eqref{opt2}. For convenience, we first reformulate $\mathcal{I}_{\!\scriptscriptstyle A}(\bs;\by_{\!\scriptscriptstyle R})$ by precoder vectorization, and then the same procedure can be applied for $\mathcal{I}_{\!\scriptscriptstyle A}(\bs;\bz_{\scriptscriptstyle E})$ and the constraints of problem \eqref{opt2}.

We start by rewriting $\be_{mk}^H\bP^H\bm{\Psi}_{\!h}\bP\be_{mk}$ as
\begin{align}\label{3B1}
\be_{mk}^H\bP^H\bm{\Psi}_{\!h}\bP\be_{mk}=\sum_{i=1}^2\be_{mk,i}^H\bP_{\!i}^H\bm{\Psi}_{\!h_{i}}\bP_{\!i}\be_{mk,i}
\end{align}
where $\be_{mk}=[\be_{mk,1}^T, \be_{mk,2}^T]^T$.
Using the following matrix equation \cite{seber2008matrix}:
\begin{align}
\trace(\bA^{\!T}\bB\bA\bC)=\mathrm{vec}(\bA)^T\cdot(\bC^T\otimes\bB)\cdot\mathrm{vec}(\bA)
\end{align}
$\be_{mk,i}^H\bP_{\!i}^H\bm{\Psi}_{\!h_{i}}\bP_{\!i}\be_{mk,i}$ can be rewritten as
\begin{align}
\be_{mk,i}^H\bP_{\!i}^H&\bm{\Psi}_{\!h_{i}}\bP_{\!i}\be_{mk,i}=\trace\big(\bP_{\!i}^H\bm{\Psi}_{\!h_{i}}\bP_{\!i}\bE_{mk,i}^T\big)\nonumber\\
&=\mathrm{vec}(\bP_{\!i})^H\cdot(\bE_{mk,i}\otimes\bm{\Psi}_{\!h_{i}})\cdot\mathrm{vec}(\bP_{\!i})
\end{align}
where $\bE_{mk,i}=(\be_{mk,i}\be_{mk,i}^H)^T$. By letting
\begin{align}
\hat{\bp}=
\begin{bmatrix}
\mathrm{vec}(\bP_{\!1})\\
\mathrm{vec}(\bP_{\!2})
\end{bmatrix},\quad
\bp=
\begin{bmatrix}
\Re\{\hat{\bp}\}\\
\Im\{\hat{\bp}\}
\end{bmatrix}
\end{align}
and
\begin{align}
&\hat{\bA}_{mk}=\frac{1}{2}\cdot \diag \big(\bE_{mk,1}\otimes\bm{\Psi}_{\!h_{1}},\bE_{mk,2}\otimes\bm{\Psi}_{\!h_{2}}\big)\\
&\bA_{mk}=
\begin{bmatrix}
\Re\{\hat{\bA}_{mk}\} &-\Im\{\hat{\bA}_{mk}\}\\
\Im\{\hat{\bA}_{mk}\} &\;\;\;\Re\{\hat{\bA}_{mk}\}
\end{bmatrix}
\end{align}
$\mathcal{I}_{\!\scriptscriptstyle A}(\bs;\by_{\!\scriptscriptstyle R})$ can be expressed alternatively as
\begin{align}
\mathcal{I}_{\!\scriptscriptstyle A}(\bs;\by_{\!\scriptscriptstyle R})=&\log N-\frac{1}{N}\sum_{m=1}^{N}\log\sum_{k=1}^{N}\nonumber\\ &\prod_{q}\Big(1+\frac{h_q}{\sigma_{\!\scriptscriptstyle R}^2}\cdot \bp^{T}\bA_{mk}\bp\Big)^{-1}.
\end{align}
Here $\bA_{mk}\succeq \mathbf{0}$ because $\bp^T\bA_{mk}\bp$ is equal to $\|\bm{\Psi}_{\!h}^{\frac{1}{2}}\bP\be_{mk}\|^2$, which is non-negative.

Similarly, we define $\hat{\bB}_{mk}$ and $\bB_{mk}$ as
\begin{align}
&\hat{\bB}_{mk}=\frac{1}{2}\cdot \diag \big(\bE_{mk,1}\otimes\bm{\Psi}_{\!g_{1}},\bE_{mk,2}\otimes\bm{\Psi}_{\!g_{2}}\big)\\
&\bB_{mk}=
\begin{bmatrix}
\Re\{\hat{\bB}_{mk}\}&-\Im\{\hat{\bB}_{mk}\}\\
\Im\{\hat{\bB}_{mk}\} &\;\;\;\Re\{\hat{\bB}_{mk}\}
\end{bmatrix}\succeq \bm{0}
\end{align}
$\hat{\bC}_{i}$ and $\bC_{i}$ as
\begin{align}
&\hat{\bC}_{i}=\diag \big(\bI \otimes (2-i)\bI, \bI\otimes(i-1)\bI\big)\\
&\bC_{i}=
\begin{bmatrix}
\Re\{\hat{\bC}_{i}\}&-\Im\{\hat{\bC}_{i}\}\\
\Im\{\hat{\bC}_{i}\} &\;\;\;\Re\{\hat{\bC}_{i}\}
\end{bmatrix}\succeq \bm{0}
\end{align}
$\hat{\bD}_{j}$ and $\bD_{j}$ as
\begin{align}
&\hat{\bD}_{j}=\trace(\bm{\Phi}_{\!f_j})\cdot \diag \big(\bI\otimes\bm{\Psi}_{\!f_{1,j}},\bI\otimes\bm{\Psi}_{\!f_{2,j}}\big)\\
&\bD_{j}=
\begin{bmatrix}
\Re\{\hat{\bD}_{j}\}&-\Im\{\hat{\bD}_{j}\}\\
\Im\{\hat{\bD}_{j}\} &\;\;\;\Re\{\hat{\bD}_{j}\}
\end{bmatrix}\succeq \bm{0}.
\end{align}
Then problem \eqref{opt2} is converted into a vectorized form
\begin{equation}\label{opt3}
\begin{aligned}
& \underset{\bp\in \mathcal{P}}{\mathrm{maximize}}
& & \big[f(\bp)-g(\bp)\big]^{+}
\end{aligned}
\end{equation}
where $f(\bp)$ and $g(\bp)$ are given below
\begin{align}
&f(\bp)=\frac{1}{N}\sum_{m=1}^{N} \log\sum_{k=1}^{N}\prod_{q}\Big(1+\frac{g_q}{\sigma_{\!\scriptscriptstyle E}^2}\cdot \bp^{T}\bB_{mk}\bp\Big)^{-1}\\
&g(\bp)=\frac{1}{N}\sum_{m=1}^{N} \log\sum_{k=1}^{N}\prod_{q}\Big(1+\frac{h_q}{\sigma_{\!\scriptscriptstyle R}^2}\cdot \bp^{T}\bA_{mk}\bp\Big)^{-1}
\end{align}
and $\mathcal{P}$ is the feasible set
\begin{align}
\mathcal{P}=\Big\{\bp \big|\bp^T\bC_{i}\bp\leq\beta_{i},i=1,2, \bp^T\bD_{j}\bp\leq\gamma_{j},\forall{j}\Big\}.
\end{align}

The feasible set $\mathcal{P}$ is convex and compact because it can be interpreted geometrically as the intersection of multiple ellipsoids. The objective function $\big[f(\bp)-g(\bp)\big]^{+}$ is continuous over $\mathcal{P}$ because both $f(\bp)$ and $g(\bp)$ are continuous functions. Therefore, the existence of a globally optimal solution is guaranteed by the Weierstrass extreme value theorem \cite{rudin1964principles}. In addition, the operator $[\cdot]^{+}$ has no effect on the optimal value of problem \eqref{opt3} and thus can be removed from the objective function because $\bp=\bm{0}$ always belongs to $\mathcal{P}$. However, it is extremely difficult to solve problem \eqref{opt3} due to the following reasons: First, both $f(\bp)$ and $g(\bp)$ are neither convex nor concave, thus \eqref{opt3} is a purely non-convex optimization problem. Second, problem \eqref{opt3} is a NP-hard problem because a specialized problem with particular
parameters $\bA_{mk}$ and $\bB_{mk}$ is NP-hard \cite{nemirovski1999maximization}.

Although $f(\bp)-g(\bp)$ is non-concave, it can be expressed as a DC function by adding a convex term
\begin{align}
\sigma(\bp)=k\cdot\bp^{T}\bp, \; k>0.
\end{align}
We can prove that both $f(\bp)+\sigma(\bp)$ and $g(\bp)+\sigma(\bp)$ are convex functions if
\begin{align}
k\geq \alpha\cdot\max\big(\trace(\bm{\Phi}_{\!h})\!\cdot\!\lambda_{\mathrm{max}}(\bm{\Psi}_{\!h}), \trace(\bm{\Phi}_{\!g})\!\cdot\! \lambda_{\mathrm{max}}(\bm{\Psi}_{\!g}) \big)
\end{align}
where $\alpha=\sum_{m,k}\|\be_{mk}\|^{2}$, $\lambda_{\mathrm{max}}(\cdot)$ represents the maximum eigenvalue of a matrix. Then $[f(\bp)+\sigma(\bp)]-[g(\bp)+\sigma(\bp)]$ is an explicit DC function, and problem \eqref{opt3} can be solved by DC algorithms. However, this DC representation is not efficient because $k$ is too large \cite{ferrer2009improving}. Through extensive simulations, we observe that even when each node in the system is only equipped with two antennas, the DC algorithm with this representation cannot converge within hundreds of thousands of iterations. Therefore, a computationally efficient DC representation of the approximated average secrecy sum rate is crucial for designing our algorithm.

\subsection{Outer Approximation of the Feasible Set}
We first rewrite \eqref{opt3} with an additional hyperrectangle $\mathcal{B}_{\mathrm{init}}$
\begin{equation}\label{opt5}
\begin{aligned}
& \underset{\bp\in \mathcal{P}\cap \mathcal{B}_{\mathrm{init}}}{\mathrm{maximize}}
& & f(\bp)-g(\bp)
\end{aligned}
\end{equation}
in which the hyperrectangle $\mathcal{B}_{\mathrm{init}}$ is given by
\begin{align}
\mathcal{B}_{\mathrm{init}}=\Big\{\bp \big| \mathbf{l}(\mathcal{B}_{\mathrm{init}})\leq\bp\leq\bu(\mathcal{B}_{\mathrm{init}})\Big\}.
\end{align}
To ensure that problems \eqref{opt5} and \eqref{opt3} are equivalent, the hyperrectangle $\mathcal{B}_{\mathrm{init}}$ should contain the feasible set $\mathcal{P}$, i.e., $\mathcal{P}\subseteq \mathcal{B}_{\mathrm{init}}$.
Let $u_{i}$ and $l_{i}$ denote the $i$-th component of $\bu(\mathcal{B}_{\mathrm{init}})$ and $\mathbf{l}(\mathcal{B}_{\mathrm{init}})$, respectively. $\mathcal{B}_{\mathrm{init}}$ can be obtained via solving the following concave maximization problem
\begin{align} \label{dbound}
u_{i}=\underset{\bp\in \mathcal{P}}{\mathrm{maximize}} \quad p_{i}
\end{align}
where $p_i$ is the $i$-th component of $\bp$. Due to the symmetry of problem \eqref{dbound}, $l_i$ can be set as $-u_i$.

By introducing a new variable $\bQ=\bp\bp^T$, we define a set function $\varphi(\mathcal{F})$ as the optimal value of the following optimization problem
\begin{equation}\label{opt6}
\begin{aligned}
\varphi(\mathcal{F})\triangleq \;& \underset{(\bQ,\bp)\in \mathcal{F}}{\mathrm{maximize}}
& & F(\bQ)-G(\bQ)
\end{aligned}
\end{equation}
where $F(\bQ)$ and $G(\bQ)$ are given below
\begin{align}
&F(\bQ)=\frac{1}{N}\sum_{m=1}^{N} \log\sum_{k=1}^{N}\prod_q\Big(1+\frac{g_q}{\sigma_{\!\scriptscriptstyle E}^2}\cdot \trace(\bB_{mk}\bQ)\Big)^{-1}\\
&G(\bQ)=\frac{1}{N}\sum_{m=1}^{N} \log\sum_{k=1}^{N}\prod_q\Big(1+\frac{h_q}{\sigma_{\!\scriptscriptstyle R}^2}\cdot \trace(\bA_{mk}\bQ)\Big)^{-1}.
\end{align}
Note that $F(\bQ)$ and $G(\bQ)$ are convex functions because 1) $\log\sum_k\!\prod_q\!f_{q,k}^{-1}$ can be written as $\log\!\sum_k\!\exp(\!-\!\sum_q \ln\!f_{q,k})$; 2) $\log\sum_{k}\exp(g_k)$ is convex whenever $g_k$ are convex \cite{boyd2004convex}. Therefore, $F(\bQ)-G(\bQ)$ is a DC function. Furthermore, when $\mathcal{F}_{\mathrm{init}}$, given by
\begin{align}
&\mathcal{F}_{\mathrm{init}}\!=\!\left\{\!(\bQ,\bp)\!\left|
\begin{aligned}
&\bQ=\bp\bp^T, \trace(\bD_{j}\bQ)\leq\gamma_{j},\; \forall j,\\
&\bp\in \mathcal{B}_{\mathrm{init}},\trace(\bC_{i}\bQ)\leq\beta_{i}, i=1,2
\end{aligned}
\right.\right\}
\end{align}
is equivalent to the feasible set $\mathcal{P}$, $\varphi(\mathcal{F}_{\mathrm{init}})$ serves as the optimal value of problem \eqref{opt5}. However, it is very difficult to obtain $\varphi(\mathcal{F}_{\mathrm{init}})$ directly because $\mathcal{F}_{\mathrm{init}}$ is a non-convex set. Although we can use semidefinite relaxation (SDR) to relax $\mathcal{F}_{\mathrm{init}}$ into a convex set by relaxing the non-convex part $\bQ\!=\!\bp\bp^T$, the solution obtained by SDR is not optimal and cannot be improved iteratively. Hence  we need tighter relaxations to overcome the shortcomings of SDR.

The key idea of our proposed precoding algorithm is to generate a sequence of asymptotically tight sets $\{\mathcal{F}_k\}$ to approach $\mathcal{F}_{\mathrm{init}}$, and then $\varphi(\mathcal{F}_{\mathrm{init}})$ can be approached iteratively from above by solving a sequence of optimization problems $\{\varphi(\mathcal{F}_k)\}$. The sequence $\{\mathcal{F}_k\}$ should satisfy the following three properties:
\begin{align}\label{OuterPro}
&\mathcal{F}_1\supseteq\mathcal{F}_2\supseteq...\supseteq \mathcal{F}_{\mathrm{init}}\nonumber\\
&\lim_{k\rightarrow\infty} \varphi(\mathcal{F}_k)=\varphi(\mathcal{F}_{\mathrm{init}})\nonumber\\
\vspace{-1.8cm}
&\mathcal{F}_k=\bigcup_{i=1}^{k}\mathcal{C}(\mathcal{B}_i), \;\forall k
\end{align}
where $\mathcal{C}(\mathcal{B}_i)$ is a convex set to be defined in \eqref{Set}. The first property implies that $\{\varphi(\mathcal{F}_k)\}$ is a monotonically decreasing sequence bounded below by $\varphi(\mathcal{F}_{\mathrm{init}})$. The second property guarantees that $\varphi(\mathcal{F}_{\mathrm{init}})$ can be readily obtained by the sequence $\{\varphi(\mathcal{F}_k)\}$. The last property provides a trackable way to compute $\{\varphi(\mathcal{F}_k)\}$, that is,
\begin{align}
\varphi(\mathcal{F}_k)=\max_{1\leq i\leq k} \varphi(\mathcal{C}(\mathcal{B}_i)).
\end{align}

Based on \eqref{OuterPro}, achieving $\varphi(\mathcal{F}_{\mathrm{init}})$ may need a sufficiently large number of iterations, which is not practical when the computational time is concerned. In order to address this issue, we also generate a lower bound of $\varphi(\mathcal{F}_{\mathrm{init}})$ in each iteration. Denote the optimal solution for $\varphi(\mathcal{F}_k)$ at the $k$-th iteration by $(\bQ_{k}^{\mathrm{opt}},\bp_{k}^{\mathrm{opt}})$. We extract a feasible solution of problem \eqref{opt5} from $\bQ_{k}^{\mathrm{opt}}$, and the corresponding approximated average secrecy sum rate is denoted by $\varphi_{\!\scriptscriptstyle L}(\mathcal{F}_k)$, which serves as a lower bound of $\varphi(\mathcal{F}_{\mathrm{init}})$.

In the remaining part of this subsection, we construct $\{\mathcal{F}_k\}$ explicitly as the union of convex sets $\{\mathcal{C}(\mathcal{B}_i)\}$. The approximated average secrecy sum rate maximization problem over $\mathcal{C}(\mathcal{B}_i)$ and an efficient method to generate the lower bound $\varphi_{\!\scriptscriptstyle L}(\mathcal{F}_k)$ are investigated in the next subsection.

For ease of exposition, we first define a convex set $\mathcal{C}(\mathcal{B})$ as
\begin{align}\label{Set}
&\mathcal{C}(\mathcal{B})\!\triangleq\!\left\{\!(\bQ,\bp)\!\left|
\begin{aligned}
&\bQ\succeq\bp\bp^T,\trace(\bC_{i}\bQ)\leq\beta_{i}, i=1,2,\\
&(\bQ,\bp)\in \mathcal{S}(\mathcal{B}),\trace(\bD_{j}\bQ)\leq\gamma_{j}, \forall j
\end{aligned}
\right.\right\}
\end{align}
where $\mathcal{S}(\mathcal{B})$ is another convex set given by
\begin{align}
&\mathcal{S}(\mathcal{B})\!\triangleq\!\left\{\!(\bQ,\bp)\!\left|
\begin{aligned}
&\bQ\!-\!\bL_{\bp}\!-\!\bL_{\bp}^T\!+\!\mathbf{l}(\mathcal{B})\!\cdot\!\mathbf{l}(\mathcal{B})^{T}\!\geq\! \bm{0},\\
&\bQ\!-\!\bU_{\bp}\!-\!\bU_{\bp}^T\!+\!\mathbf{u}(\mathcal{B})\!\cdot\!\mathbf{u}(\mathcal{B})^T\!\geq\! \bm{0},\\
&\bQ\!-\!\bL_{\bp}\!-\!\bU_{\bp}^T\!+\!\mathbf{l}(\mathcal{B})\!\cdot\!\mathbf{u}(\mathcal{B})^T\!\geq\! \bm{0},\\
&\mathbf{l}(\mathcal{B})\leq\bp\leq \bu(\mathcal{B})
\end{aligned}
\right.\right\}
\end{align}
with $\bL_{\bp}=\mathbf{l}(\mathcal{B})\cdot\bp^T$ and $\bU_{\bp}=\mathbf{u}(\mathcal{B})\cdot\bp^T$.
The following two propositions are the foundation for constructing $\{\mathcal{F}_k\}$.
\begin{proposition}
If we split the initial hyperrectangle $\mathcal{B}_{\mathrm{init}}$ into $K$ smaller hyperrectangles such that $\mathcal{B}_{\mathrm{init}}=\mathcal{B}_1 \cup ... \cup \mathcal{B}_K$, then $\mathcal{F}_{\mathrm{init}} \subseteq\mathcal{C}(\mathcal{B}_1) \cup ... \cup \mathcal{C}(\mathcal{B}_K)$.
\end{proposition}
\begin{IEEEproof}
See Appendix A.
\end{IEEEproof}
\begin{proposition}
If we split a hyperrectangle $\mathcal{B}$ into two smaller hyperrectangles $\mathcal{B}_1$ and $\mathcal{B}_2$ such that $\mathcal{B}=\mathcal{B}_1 \cup \mathcal{B}_2$ and $\mathcal{B}_1 \cap \mathcal{B}_2=\emptyset$, then $\mathcal{C}(\mathcal{B}_1) \cup \mathcal{C}(\mathcal{B}_2) \subseteq \mathcal{C}(\mathcal{B})$.
\end{proposition}
\begin{IEEEproof}
See Appendix A.
\end{IEEEproof}

With the help of Proposition 1, the first relaxed set $\mathcal{F}_1$ is obtained
\begin{align}
\mathcal{F}_1=\mathcal{C}(\mathcal{B}_{\mathrm{init}}).
\end{align}
Similarly, in the second iteration, we generate $\mathcal{F}_2$ by partitioning the initial hyperrectangle $\mathcal{B}_{\mathrm{init}}$ into two non-intersection hyperrectangles $\mathcal{B}_1$ and $\mathcal{B}_2$
\begin{align}
\mathcal{F}_2=\mathcal{C}(\mathcal{B}_1) \cup \mathcal{C}(\mathcal{B}_2)\subseteq \mathcal{F}_1.
\end{align}

We continue this process to generate a sequence of relaxed sets $\{\mathcal{F}_k\}$ satisfying \eqref{OuterPro}. At the $k$-th iteration, $\mathcal{B}_{\mathrm{init}}$ is split into $k$ non-intersection hyperrectangles $\mathcal{B}_1, \mathcal{B}_2,...,\mathcal{B}_k$ such that
\begin{align}
\mathcal{F}_k=\mathcal{C}(\mathcal{B}_1) \cup ... \cup\mathcal{C}(\mathcal{B}_k).
\end{align}
The outer approximation algorithm is summarized in Algorithm 1.
\begin{algorithm}
\caption{: The outer approximation algorithm}
\begin{algorithmic}
\STATE 1) Initialization: Set the maximum number of iterations \STATE \quad $K_{\mathrm{max}}$, $k=1$,  $\mathbb{B}=\{\mathcal{B}_{\mathrm{init}}\}$, $\mathcal{F}_1=\mathcal{C}(\mathcal{B}_{\mathrm{init}})$, $U_1=\varphi(\mathcal{F}_1)$ \STATE\quad and $L_1=\varphi_{\!\scriptscriptstyle L}(\mathcal{F}_1)$.
\STATE 2) Stopping criterion: if $k\leq K_{\text{max}}$ go to the next step, oth-\STATE\quad erwise STOP.
\STATE 3) Partition criterion:
\STATE \quad a) select $\mathcal{B}_{g}=\mathrm{arg}\max_{\mathcal{B}\in \mathbb{B} }\big\{\varphi(\mathcal{C}(\mathcal{B}))\big\}$.
\STATE \quad b) split $\mathcal{B}_{g}$ along any of its longest edge into two small\STATE\quad\;\;\;
hyperrectangles, $\mathcal{B}_{\uppercase\expandafter{\romannumeral1}}$ and $\mathcal{B}_{\uppercase\expandafter{\romannumeral2}}$, with equal volume.
\STATE \quad c) remove $\mathcal{B}_{g}$ from $\mathbb{B}$, and add $\mathcal{B}_{I}$ and $\mathcal{B}_{II}$ into $\mathbb{B}$.
\STATE \quad d) compute the upper and lower bounds of $\varphi(\mathcal{F}_{\mathrm{init}})$
 \begin{align*}
 &\mathcal{F}_{k+1}=\bigcup_{\mathcal{B}\in \mathbb{B}} \mathcal{C}(\mathcal{B})\\
 &U_{k+1}=\varphi(\mathcal{F}_{k+1})=\max_{\mathcal{B}\in \mathbb{B}} \big\{\varphi(\mathcal{C}(\mathcal{B}))\big\}\\
 &L_{k+1}=\varphi_{\!\scriptscriptstyle L}(\mathcal{F}_{k+1}).
 \end{align*}
\STATE 4) Set $k:= k+1$ and go to step 2).
\end{algorithmic}
\end{algorithm}

The convergence of Algorithm 1 is presented by the following proposition.
\begin{proposition}
The sequence $\{\varphi(\mathcal{F}_k)\}$ converges to $\varphi(\mathcal{F}_{\mathrm{init}})$, i.e., $\forall \varepsilon>0, \exists K>0$, such that $k>K$ implies $\varphi(\mathcal{F}_{\mathrm{init}})<\varphi(\mathcal{F}_k)<\varphi(\mathcal{F}_{\mathrm{init}})+\varepsilon$.
\end{proposition}
\begin{IEEEproof}
See Appendix B.
\end{IEEEproof}

It is worth remarking that each relaxed set $\mathcal{F}_k$ is tighter than the set relaxed by SDR. We denote $\mathcal{F}_{\mathrm{sdr}}=\big\{\bQ|\bQ\succeq\bm{0},\trace(\bC_{i}\bQ)\leq\beta_{i}, i=1,2,\trace(\bD_{j}\bQ)\leq\gamma_{j}, \forall j\big\}$. Since $\bp\bp^{T}\succeq\bm{0}$, we have $\big\{\bQ|(\bQ,\bp)\in \mathcal{F}_k\big\}\subseteq\mathcal{F}_{\mathrm{sdr}}$ for any $k$. Thus the solution obtained by Algorithm 1 is better than that of the SDR method.

\subsection{DC Optimization Over the Convex Set}
In this subsection, we maximize the approximated average sum rate over the convex set $\mathcal{C}(\mathcal{B})$ by employing the convex-concave procedure \cite{yuille2003concave}. The convex-concave procedure is a general polynomial time algorithm for solving DC problems, and it works quite well in practice\cite{beck2010sequential,mehanna2015feasible,khabbazibasmenj2012sum}. We first rewrite the optimization problem as follows
\begin{equation}\label{opt7}
\begin{aligned}
\varphi(\mathcal{C}(\mathcal{B}))=\;& \underset{(\bQ,\bp)\in \mathcal{C}(\mathcal{B})}{\mathrm{maximize}}
& & F(\bQ)-G(\bQ).
\end{aligned}
\end{equation}
The objective function of problem \eqref{opt7} is a DC function, and the convex part $F(\bQ)$ can be lower bounded by its tangent at any point $\bQ_{\mathrm{c}}\succeq \bm{0}$
\begin{align}\label{3C10}
F(\bQ)\geq F(\bQ_{\mathrm{c}})+\trace\big\{\nabla F(\bQ_{\mathrm{c}})^T(\bQ-\bQ_{\mathrm{c}})\big\}
\end{align}
where $\nabla F(\bQ_{\mathrm{c}})$ is the gradient of $F(\bQ)$ at $\bQ_{\mathrm{c}}$
\begin{align}
\nabla F(\bQ_{\mathrm{c}})=-\frac{1}{N}\sum_{m,k}w_{mk}\sum_{q}\frac{g_q\!\cdot\!\bB_{mk}^{T}}{\sigma_{\!\scriptscriptstyle E}^2+g_q\!\cdot\!\trace(\bB_{mk}\bQ_{\mathrm{c}})}
\end{align}
with
\begin{align}
w_{mk}=\frac{1}{\ln(2)}\cdot\frac{\exp\big\{\sigma_{\!\scriptscriptstyle E}^2+g_q\cdot\trace(\bB_{mk}\bQ_{\mathrm{c}})\big\}}{\sum_{k}\exp\big\{\sigma_{\!\scriptscriptstyle E}^2+g_q\cdot\trace(\bB_{mk}\bQ_{\mathrm{c}})\big\}}.
\end{align}
Therefore, by replacing $F(\bQ)-G(\bQ)$ with a concave lower bound
\begin{align}\label{con-app}
\hat{F}(\bQ;\bQ_{\mathrm{c}})\!=\!F(\bQ_{\mathrm{c}})\!+\!\trace\big\{\nabla F(\bQ_{\mathrm{c}})^T(\bQ\!-\!\bQ_{\mathrm{c}})\big\}\!-\!G(\bQ)
\end{align}
we obtain the following concave maximization problem
\begin{equation}\label{opt8}
\begin{aligned}
& \underset{(\bQ,\bp)\in \mathcal{C}(\mathcal{B})}{\mathrm{maximize}}
& & \hat{F}(\bQ;\bQ_{\mathrm{c}}).
\end{aligned}
\end{equation}

The convex-concave procedure obtains a locally optimal solution of problem \eqref{opt7} by solving a sequence of concave maximization problems \eqref{opt8} with different $\bQ_\mathrm{c}$. Once the optimal solution of \eqref{opt8} in the first iteration is found at initial $\bQ_\mathrm{c}$, denoted as $\bQ^{*}_{1}$, the algorithm replaces $\bQ_\mathrm{c}$ with $\bQ^{*}_{1}$ and then solve \eqref{opt8} again. At the $n$-th iteration, the optimal solution of \eqref{opt8} is obtained by replacing $\bQ_\mathrm{c}$ with $\bQ^{*}_{n-1}$, which is the optimal solution at the $(n-1)$-th iteration. The convex-concave procedure for solving problem \eqref{opt7} is summarized in Algorithm 2.

\begin{algorithm}
\caption{: The convex-concave procedure}
\begin{algorithmic}
\STATE 1) Initialization: Given tolerance $\epsilon>0$, choose a random \STATE\quad\;initial point $\bQ_0\succeq \bm{0}$, set $N=1$,
$s_{0}=F(\bQ_0)-G(\bQ_0)$, \STATE\quad\;$s_{1}=F(\bQ^{*}_{1})-G(\bQ^{*}_{1})$. Let $\bQ^{*}_{n}$ represent the optimal \STATE\quad\;solution of \eqref{opt8} at the $n$-th iteration.
\STATE 2) Stopping criterion: if $|s_{n}-s_{n-1}|>\epsilon$ go to the next step, \STATE\quad\;otherwise STOP.
\STATE 3) Convex approximation:
\STATE \quad a) set $\bQ_{\mathrm{c}}=\bQ^{*}_{n}$ and solve problem \eqref{opt8} to obtain $\bQ^{*}_{n+1}$.
\STATE \quad b) set $s_{n+1}=F(\bQ^{*}_{n+1})-G(\bQ^{*}_{n+1})$ and $\bQ_{\mathrm{opt}}=\bQ^{*}_{n+1}$.
\STATE 4) Set $n:=n+1$ and go to step 2).
\STATE 5) Output: $\bQ_{\mathrm{opt}}$ and $s_n$.
\end{algorithmic}
\end{algorithm}
The stopping criterion in Algorithm 2 is guaranteed to be satisfied due to the following proposition.
\begin{proposition}
The sequence $\{s_n\}$ generated by Algorithm 2 is monotonically increasing, i.e., $s_{n+1}\geq s_n$.
\end{proposition}
\begin{IEEEproof}
Since the feasible set of \eqref{opt8} does not change in each iteration, the optimal solution in the $n$-th iteration $\bQ^{*}_{n}$ is a feasible point in the $(n+1)$-th iteration. Thus we have
\begin{align}
\hat{F}(\bQ^{*}_{n+1};\bQ^{*}_{n})\geq \hat{F}(\bQ^{*}_{n};\bQ^{*}_{n})=s_n.
\end{align}
According to \eqref{3C10}, it follows that
\begin{align}
s_{n+1}=\hat{F}(\bQ^{*}_{n+1};\bQ^{*}_{n+1})\geq \hat{F}(\bQ^{*}_{n+1};\bQ^{*}_{n}).
\end{align}
Therefore $\{s_n\}$ is monotonically increasing.
\end{IEEEproof}

Since problem \eqref{opt7} is non-convex, Algorithm 2 is not guaranteed to converge to the globally optimal value $\varphi(\mathcal{C}(\mathcal{B}))$. Therefore, by embedding Algorithm 2 into Algorithm 1, we obtain a near optimal solution $\bQ_{k}^{\mathrm{opt}}$ and the corresponding approximated upper bound of $\varphi(\mathcal{F}_{\mathrm{init}})$ at the $k$-th iteration of Algorithm 1. Simulation results show that the gap between the approximated upper bound and the actual upper bound is usually very small because Algorithm 2 is insensitive to the initial point $\bQ_0$.

After obtaining $\bQ_{k}^{\mathrm{opt}}$ at the $k$-th iteration, we need to get a feasible precoder pair $(\bP_1,\bP_2)$ and the corresponding  lower bound $\varphi_{\!\scriptscriptstyle L}(\mathcal{F}_k)$. The feasible precoders can be obtained by extracting a feasible solution of \eqref{opt5} from $\bQ_{k}^{\mathrm{opt}}$. There are several rank one approximation methods to do this, and we adopt the Gaussian randomization procedure \cite{luo2010semidefinite}, which is summarized in Algorithm 3.

\begin{algorithm}
\caption{: Gaussian randomization procedure}
\begin{algorithmic}
\STATE 1) Given a number of randomizations $L$, and set $l=1$.
\STATE 2) If $l\leq L$ go to the next step, otherwise STOP.
\STATE 3) Generate $\bm{\xi}_l\sim N(\bm{0},\bQ_{k}^{\mathrm{opt}})$, and construct a feasible point \STATE\quad\;$\tilde{\bp}_l$
\begin{align*}
\tilde{\bp}_l=\frac{\bm{\xi}_l}{\sqrt{\max\big\{\{\frac{\bm{\xi}_l^T\bC_i\bm{\xi}_l}{\beta_i}\}_{i=1,2}, \{\frac{\bm{\xi}_l^T\bD_j\bm{\xi}_l}{\gamma_j}\}_{\forall j}\big\}}}.
\end{align*}
\STATE 4) Set $l:=l+1$ and go to step 2).
\STATE 5) Choose $\tilde{\bp}=\arg\max_{1\leq l\leq L}\; f(\tilde{\bp}_l)-g(\tilde{\bp}_l)$.
\STATE 6) Set $\varphi_{\!\scriptscriptstyle L}(\mathcal{F}_k)=f(\tilde{\bp})-g(\tilde{\bp})$.
\STATE 7) Recover $(\bP_{\!1},\bP_{\!2})$ from $\tilde{\bp}$.
\end{algorithmic}
\end{algorithm}

\subsection{Complexity Analysis}
The computational complexity of Algorithm 1 is analyzed as follows. In each iteration, Algorithm 1 invokes Algorithms 2 and 3 twice to calculate the approximated upper bound and the lower bound. Since the complexity of Algorithm 3 is negligible, the complexity order for Algorithm 1 is given by
\begin{align}\label{CO}
2K_{\mathrm{max}}\cdot C
\end{align}
where $K_{\mathrm{max}}$ is the maximum number of iterations, and $C$ is the complexity order for Algorithm 2. Algorithm 2 obtains a local maxima of problem \eqref{opt7} by solving a sequence of concave maximization problems \eqref{opt8}. Each concave maximization problem \eqref{opt8} can be solved by the interior point method, and the complexity order is about $\mathcal{O}(N^3)$ \cite{boyd2004convex}, where $N=4(N_{T_1}^2+N_{T_2}^2)^2+2(N_{T_1}^2+N_{T_2}^2)$ is the total number of optimization variables in problem \eqref{opt8}. Assuming that Algorithm 2 solves problems \eqref{opt8} $T$ times, the complexity order for Algorithm 2 is given by $\mathcal{O}(T\!\cdot\! N^3)$. Based on \eqref{CO}, the overall complexity order for Algorithm 1 is then $\mathcal{O}(2K_{\mathrm{max}}T\!\cdot\! N^3)$.

\section{Numerical Results}
In this section, we provide numerical results to demonstrate the efficacy of our proposed algorithm for the fading CMAC-WT under finite-alphabet inputs. For illustration purpose, we adopt the exponential correlation model:
\begin{align} \label{CorrM}
[\bC(\rho)]_{i,j}=\rho^{|i-j|}, \quad \forall (i,j)
\end{align}
where the scalar $\rho\in [0,1)$ depicts the interference coupling between different antennas.

\subsection{Convergence and Complexity Analysis}
The convergence behavior of the proposed algorithm is demonstrated by considering a two-user fading CMAC-WT with two STs, one SR, one ED, and one PR. Each node in the system has two antennas. The correlation matrices are given by
\begin{align}
&\bm{\Phi}_{\!h}=\bC(0.3), \bm{\Psi}_{\!h_{1}}=\bC(0.95), \bm{\Psi}_{\!h_{2}}=\bC(0.85)\nonumber\\ &\bm{\Phi}_{\!g}=\bC(0.6), \bm{\Psi}_{\!g_{1}}=\bC(0.4), \bm{\Psi}_{\!g_{2}}=\bC(0.95)\nonumber\\
&\bm{\Phi}_{\!f}=\bC(0.5), \bm{\Psi}_{\!f_{1}}=\bC(0.3), \bm{\Psi}_{\!f_{2}}=\bC(0.5).
\end{align}

The maximum transmit power is constrained by $\beta_1=\beta_2=2$. The interference threshold is given as $\gamma=0.2$. The input data vectors $\bs_1$ and $\bs_2$ are drawn independently from BPSK constellation, and the noise power is set as $\sigma^2_{\!\scriptscriptstyle R}=\sigma^2_{\scriptscriptstyle E}=0.1$.
\begin{figure}[!htbp]
\vspace{-0.5cm}
\centering
  \includegraphics[scale=0.33]{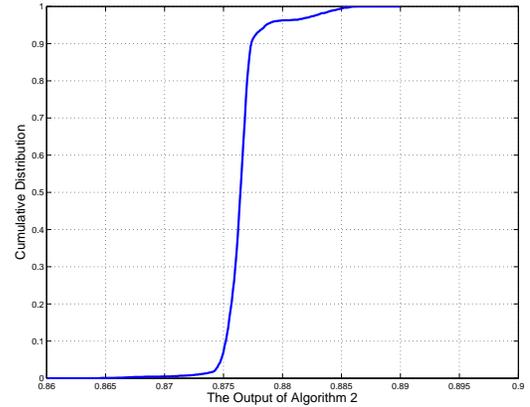}\\
  \caption{Empirical cumulative distribution of the output $s_n$ of Algorithm 2 from 3000 random initial points.}
  \vspace{-0.5cm}
\end{figure}

The empirical cumulative distribution of the output $s_n$ of Algorithm 2 from 3000 random initial points is shown in Fig. 2. The tolerance $\varepsilon$ in Algorithm 2 is set as $0.002$. $\mathbf{l}(\mathcal{B})$ and $\bu(\mathcal{B})$ are given as $\mathbf{l}(\mathcal{B})=-\sqrt{2}\!\cdot\! \bm{1}$, $\mathbf{u}(\mathcal{B})=\sqrt{2}\!\cdot\! \bm{1}$. The empirical cumulative distribution illustrates that Algorithm 2 is insensitive to the initial point. Therefore, although problem \eqref{opt7} is non-convex, the approximated upper bound obtained by Algorithm 2 is accurate.

\begin{figure}[!htbp]
\vspace{-0.45cm}
\centering
  \includegraphics[scale=0.34]{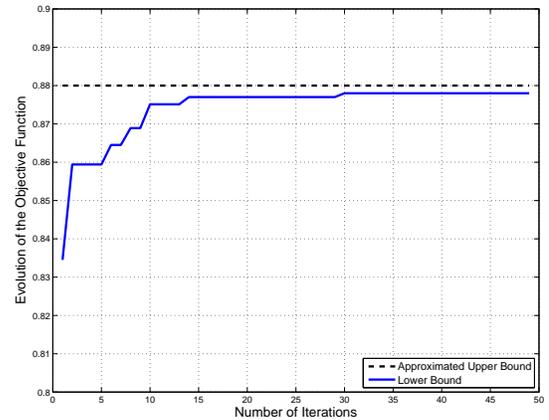}\\
  \caption{Evolution of the objective function in \eqref{opt5} with BPSK inputs.}
  \vspace{-0.15cm}
\end{figure}

Fig. 3 illustrates the evolution of the approximated upper bound and the lower bound of $\varphi(\mathcal{F}_{\mathrm{init}})$. In order to guarantee that the approximated upper bound is accurate enough, the tolerance $\varepsilon$ in Algorithm 2 is set as $0.001$. In each iteration of Algorithm 1, we invoke Algorithm 2 to generate the approximated upper bound, which can be seen as the actual upper bound of $\varphi(\mathcal{F}_{\mathrm{init}})$ according to the result in Fig. 2.
\begin{figure*}
\centering
\begin{minipage}[t]{0.44\textwidth}
\centering
  \includegraphics[scale=0.49]{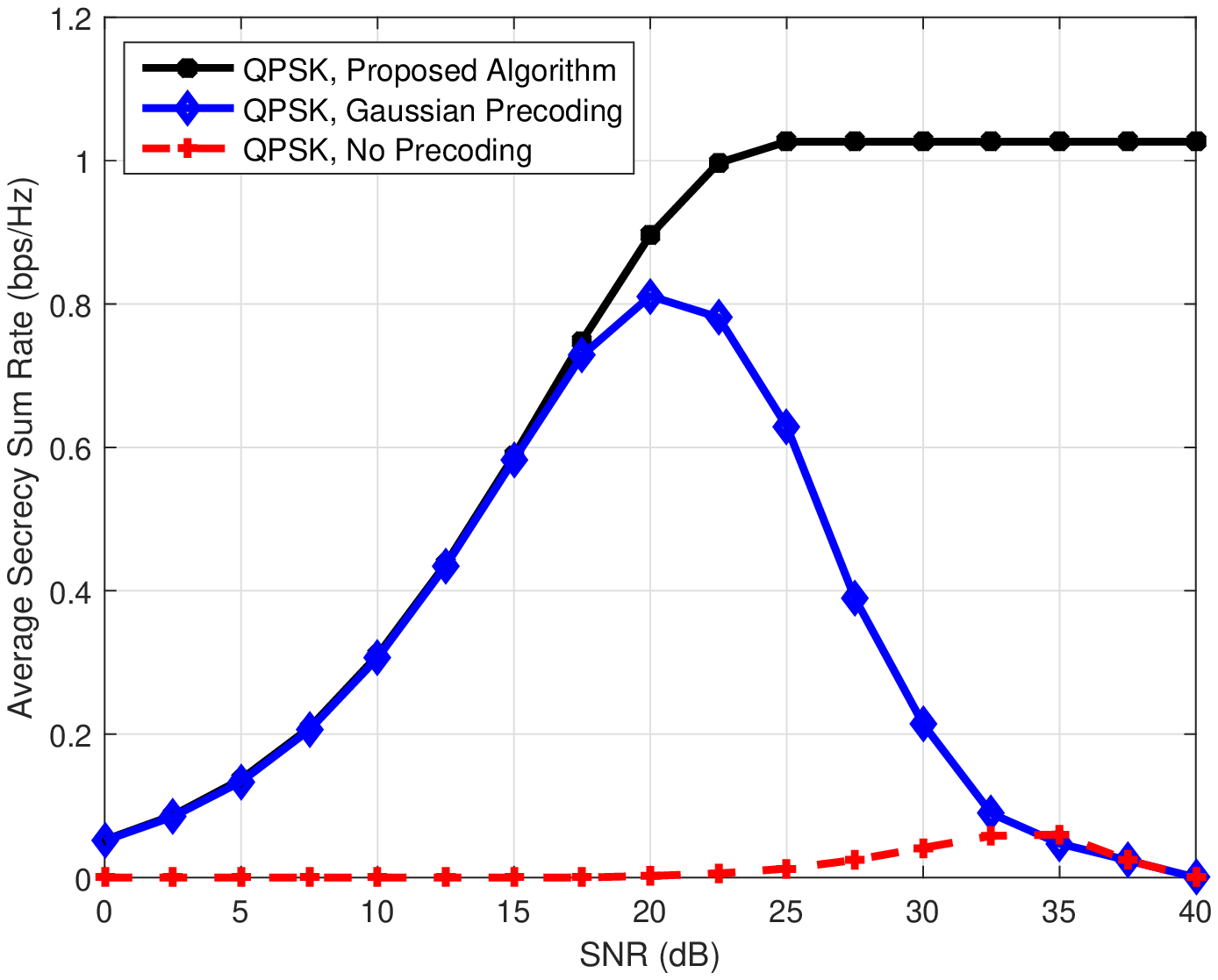}
  \caption{The interference threshold at the PR is 10 dB less than the transmit \\power ($\gamma_1=0.2$).}
\end{minipage}
\begin{minipage}[t]{0.44\textwidth}
\centering
  \includegraphics[scale=0.49]{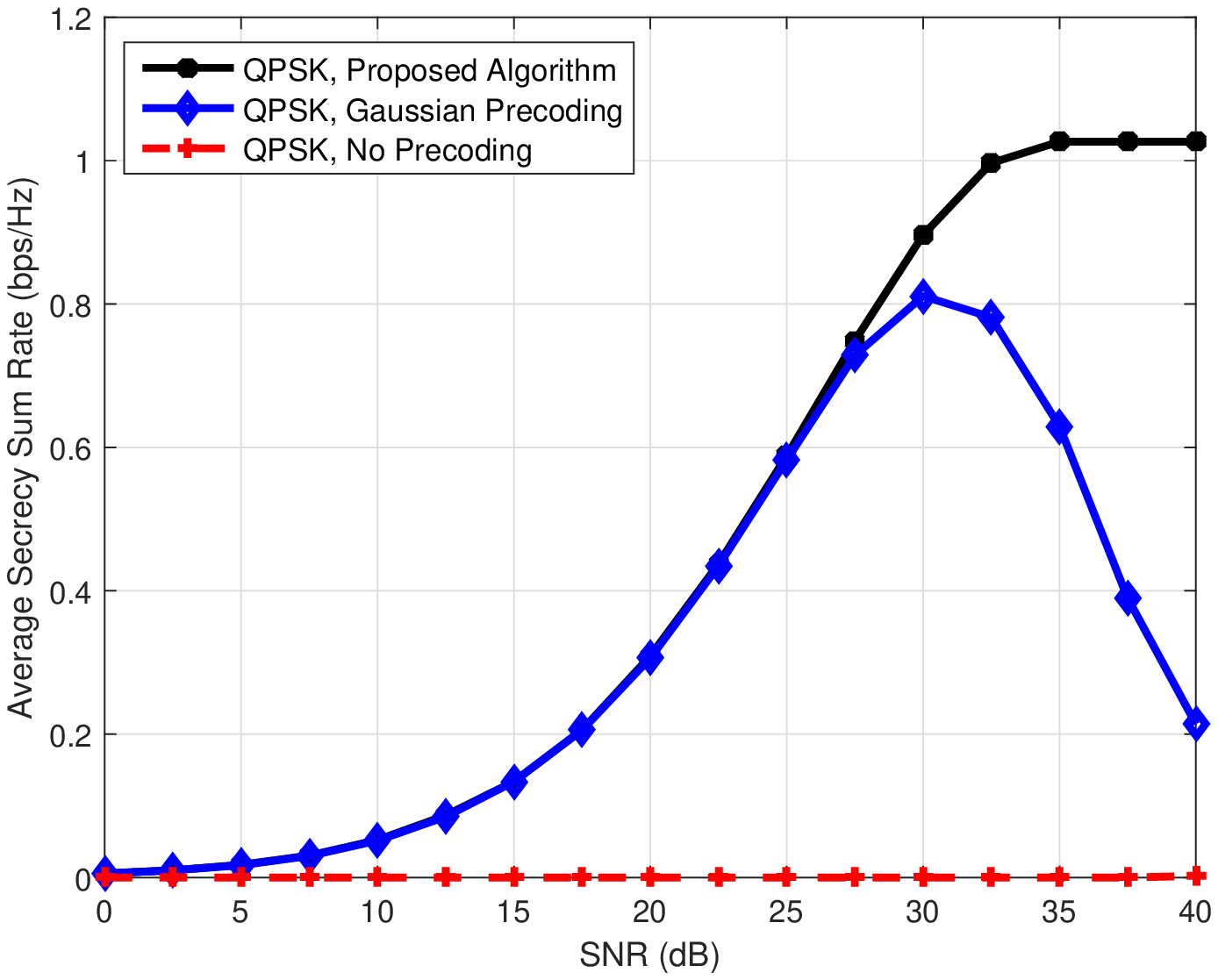}
  \caption{The interference threshold at the PR is 20 dB less than the transmit \\power ($\gamma_1=0.02$).}
\end{minipage}
\vspace{-0.44cm}
\end{figure*}
We also invoke Algorithm 3 to generate feasible precoding matrices and the corresponding lower bound $\varphi_{\!\scriptscriptstyle L}(\mathcal{F}_{\mathrm{init}})$.
Note that when all hyperrectangles in $\mathbb{B}$ shrink down to a point, we can ensure that the approximated upper bound serves exactly as the actual upper bound. From the figure, we can see that after 10 iterations, the gap between the approximated upper bound and the lower bound is less than 0.005. Moreover, near-optimal precoders within 0.002 tolerance is obtained through Algorithm 1 after 30 iterations.

\subsection{Comparison with Other Possible Methods}
In this subsection, we consider a secure cognitive radio system that has two STs, one SR, one ED, and one PR. Each node in the system has two antennas. The correlation matrices are given by
\begin{align}
&\bm{\Phi}_{\!h}=\bC(0.25), \bm{\Psi}_{\! h_{1}}=\bC(0.95), \bm{\Psi}_{\!h_{2}}=\bC(0.9)\nonumber\\
&\bm{\Phi}_{\!g}=\bC(0.75), \bm{\Psi}_{\!g_{1}}=\bC(0.5), \bm{\Psi}_{\!g_{2}}=\bC(0.3)\nonumber\\
&\bm{\Phi}_{\!f}=\bC(0.5), \bm{\Psi}_{\!f_{1}}=\bC(0.8), \bm{\Psi}_{\!f_{2}}=\bC(0.5).
\end{align}
The transmit power constraint is set as $\beta_1=\beta_2=\beta=2$. The interference thresholds $\gamma_1=0.2$ and $\gamma_1=0.02$ are considered. The modulation is QPSK and the noise variance $\sigma_{\!\scriptscriptstyle R}^2=\sigma_{\!\scriptscriptstyle E}^2=\sigma^2$. Then the SNR can be defined as $\textrm{SNR}=\beta/\sigma^2$.

Figs. 4 and 5 depict the comparison results among the Gaussian precoding method and no precoding case. The Gaussian precoding method is to design transmit covariance matrices that maximize the average secrecy sum rate under Gaussian signaling, i.e.,
\begin{equation}\label{GP}
\begin{aligned}
& \underset{\bQ_1,\bQ_2}{\mathrm{maximize}}
& & E_{\scriptscriptstyle\bH_1,\bH_2}(R_1)-E_{\scriptscriptstyle\bG_1,\bG_2}(R_2)\\
& \mathrm{subject \;to}
& & \trace(\bQ_{i})\leq\beta_{i}, \; i=1,2\\
&&& \trace(\bm{\Phi}_{\!f_j})\!\cdot\!\trace(\bQ_1\bm{\Psi}_{\!f_{1,j}}\!+\!\bQ_2\bm{\Psi}_{\!f_{2,j}})\leq\gamma_{j}, \; \forall{j}
\end{aligned}
\end{equation}
where $\bQ_i$ is the transmit covariance matrix of the $i$-th ST, $i=1,2$; $R_1$ and $R_2$ are given by
\vspace{-0.07cm}
\begin{align}
&R_1=\log\det(\bI+\frac{1}{\sigma_{\!\scriptscriptstyle R}^{2}}\bH_{1}\bQ_{1}\bH_{1}^{H}+\frac{1}{\sigma_{\!\scriptscriptstyle R}^{2}}\bH_{2}\bQ_{2}\bH_{2}^{H})\\
&R_2=\log\det(\bI+\frac{1}{\sigma_{\!\scriptscriptstyle E}^{2}}\bG_{1}\bQ_{1}\bG_{1}^{H}+\frac{1}{\sigma_{\!\scriptscriptstyle E}^{2}}\bG_{2}\bQ_{2}\bG_{2}^{H}).
\end{align}
\vspace{-0.15cm}

Problem \eqref{GP} is a DC optimization problem, thus it can be solved efficiently by DC algorithms proposed in \cite{khabbazibasmenj2012sum}. After obtaining the optimal transmit covariance matrices $(\bar{\bQ}_1,\bar{\bQ}_2)$, we can evaluate the finite-alphabet based average secrecy sum rate under the corresponding optimal precoders $(\bar{\bQ}_1^{\frac{1}{2}},\bar{\bQ}_2^{\frac{1}{2}})$. In the no precoding case, we set precoding matrices as $\bP_{\!i}=\frac{\beta_i}{N_{T_i}}\bI$, $i=1,2$, and then scale them down to meet interference threshold constraints:
\begin{align}
\bar{\bP}_{\!i}\!=\!\bigg[\max_{1\leq j\leq J}\bigg\{\frac{\trace(\bm{\Phi}_{\!f_j})}{\gamma_{j}}\!\cdot\! \sum_{i=1}^2 \trace\big(\bP_{\!i}^{H}\bm{\Psi}_{\!f_{i,j}}\bP_{\!i}\big)\bigg\}\bigg]^{-\frac{1}{2}}\!\cdot\!\bP_{\!i}.
\end{align}

Based on the results in Figs. 4 and 5, we have the following remarks:

1) In the low SNR regime, our proposed precoding algorithm and the Gaussian precoding method have the same performance. According to \cite{perez2010mimo}, the low-SNR expansion of the mutual information is irrelevant to the input distribution, thus the optimal precoders designed under Gaussian inputs are also optimal for finite-alphabet inputs case.

2) In the medium and high SNR regime, our proposed precoding algorithm offers much higher average secrecy sum rate than the Gaussian precoding method. In Fig. 4, the normalized optimal precoders designed by our proposed precoding algorithm in the high SNR regime are given by
\begin{align}
&\frac{1}{\sigma}\bP^{\mathrm{opt}}_{\!1}=
\begin{bmatrix}
0.663 + 0.008i &-1.188 + 0.277i\\
0.663 + 0.008i &-1.188 + 0.277i
\end{bmatrix}\label{T1T}\\
&\frac{1}{\sigma}\bP^{\mathrm{opt}}_{\!2}=
\begin{bmatrix}
-0.578 + 0.399i &1.209 - 0.459i\\
-0.578 + 0.399i &1.209 - 0.459i
\end{bmatrix}\label{T2T}.
\end{align}
Equations \eqref{T1T} and \eqref{T2T} imply that when the noise power $\sigma^2$ is decreased, we should reduce the optimal transmit power $\trace\big((\bP^{\mathrm{opt}}_{\!1})^H\bP^{\mathrm{opt}}_{\!1}\big)$ and $\trace\big((\bP^{\mathrm{opt}}_{\!2})^H\bP^{\mathrm{opt}}_{\!2}\big)$ such that the average secrecy sum rate is kept at the maximum value 1.0265 bpz/Hz in the high SNR regime. Furthermore, the performance of the Gaussian precoding method degrades severely with the increasing SNR in the high SNR regime. The reason is that both $E_{\scriptscriptstyle\bH}\mathcal{I}(\bs;\by_{\!\scriptscriptstyle R})$ and $E_{\scriptscriptstyle\bG}\mathcal{I}(\bs;\bz_{\scriptscriptstyle E})$ in \eqref{EMI1} and \eqref{EMI2} will saturate at $\log N$ in the high SNR regime. Therefore, if we do not carefully control the transmit power, the average secrecy sum rate with finite-alphabet inputs $E_{\scriptscriptstyle\bH}\mathcal{I}(\bs;\by_{\!\scriptscriptstyle R})-E_{\scriptscriptstyle\bG}\mathcal{I}(\bs;\bz_{\scriptscriptstyle E})$ approaches zero in the high SNR regime. Since the Gaussian precoding method ignores the saturation property of finite-alphabet inputs systems, the corresponding average secrecy sum rate with finite-alphabet inputs degrades severely in the high SNR regime.

3) Since the average secrecy sum rate for the Gaussian precoding method decreases with the increasing SNR in the high SNR regime, we can use a portion of the available transmit power to make sure that the SNR is maintained at a certain level. The average secrecy sum rate is then kept at its maximum value. This simple power control method has been used in \cite{geraci2013large,yang2014confidential} to improve the secrecy sum-rate performance at the high SNR regime.

4) The interference threshold constraints have a huge impact on the system performance. For example, when
SNR is 20 dB, the average secrecy sum rate is 0.90 bps/Hz and 0.31 bps/Hz for $\gamma_1=0.2$ and $\gamma_1=0.02$, respectively. More specifically, given the set of all feasible precoding matrices
\begin{align}
&\mathcal{P}\!=\!\left\{\!(\bP_{\!1},\bP_{\!2})\!\left|
\begin{aligned}
&\trace\big(\bP_{\!i}^H\bP_{\!i}\big)\leq\beta_{i},i=1,2,\\
&\trace(\bm{\Phi}_{\!f_j})\!\cdot\!\sum_{i=1}^2 \trace\big(\bP_{\!i}^H\mathbf{\Psi}_{\!f_{i,j}}\bP_{\!i}
\big)\!\leq\! \gamma_{j}, \forall j
\end{aligned}
\right.\!\right\}
\end{align}
we define the following parameters
\begin{align}
\bar{\beta}_i=\min \bigg\{\min_{1\leq j\leq J} \Big\{\frac{\gamma_{j}}{\trace(\bm{\Phi}_{\!f_j})\!\cdot\! \lambda_{\mathrm{min}}(\mathbf{\Psi}_{\!f_{i,j}})}\Big\},\beta_i\bigg\}
\end{align}
where $\lambda_{\mathrm{min}}(\bA)$ represents the smallest eigenvalue of $\bA$. Then for all $(\bP_{\!1},\bP_{\!2})\in \mathcal{P}$, we can easily prove that
\begin{align}\label{CCC}
\trace\big(\bP_{\!i}^H\bP_{\!i}\big)\leq \bar{\beta}_i, \quad i=1,2.
\end{align}
Equation \eqref{CCC} implies that when $\bar{\beta}_i<\beta_i$, $i=1,2$, the power constraints in $\mathcal{P}$ are inactive, i.e., only a portion of the available transmit power can be used in order to meet all interference threshold constraints. In the case of Figs. 4 and 5, $(\bar{\beta}_1,\bar{\beta}_2)$ is calculated as
\begin{align}
(\bar{\beta}_1,\bar{\beta}_2)=\left\{
\begin{aligned}
&(0.5,0.2) \quad \gamma_1=0.2\\
&(0.05,0.02) \quad \gamma_1=0.02
\end{aligned}
\right.
\end{align}
Since $(\beta_1,\beta_2)=(2,2)$, the sum rate performance in Figs. 4 and 5 is only constrained by interference threshold constraints.

5) The performance of no precoding case is very poor because we do not exploit any statistical CSI from STs to the SR and the ED.

\subsection{Comparison of Different Modulations}
Finally, we investigate the average secrecy sum rate with different modulations. We consider a secure cognitive radio system with two STs, one SR, one ED and two PRs. Each node is equipped with two antennas. The correlation matrices are given by
\begin{align}
&\bm{\Phi}_{\!h}=\bC(0.3), \bm{\Psi}_{\! h_{1}}=\bC(0.9), \bm{\Psi}_{\!h_{2}}=\bC(0.95)\nonumber\\
&\bm{\Phi}_{\!g}=\bC(0.6), \bm{\Psi}_{\!g_{1}}=\bC(0.7), \bm{\Psi}_{\!g_{2}}=\bC(0.2)\nonumber\\
&\bm{\Phi}_{\!f_1}=\bC(0.4), \bm{\Psi}_{\!f_{1,1}}=\bC(0.6), \bm{\Psi}_{\!f_{2,1}}=\bC(0.4)\nonumber\\
&\bm{\Phi}_{\!f_2}=\bC(0.5), \bm{\Psi}_{\!f_{1,2}}=\bC(0.3), \bm{\Psi}_{\!f_{2,2}}=\bC(0.5).
\end{align}
\begin{figure}[h]
  \begin{center}
  \includegraphics[scale=0.5]{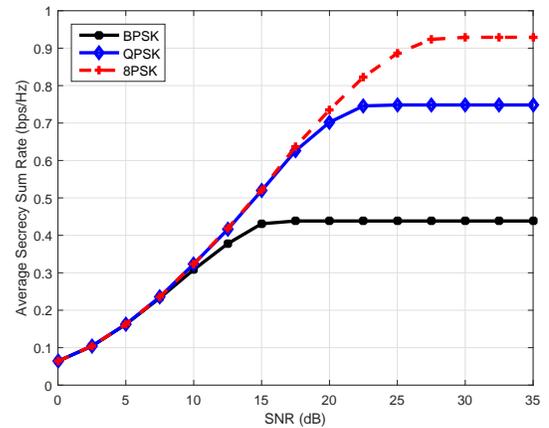}\\
  \caption{Average secrecy sum rate for the fading CMAC-WT with different modulations.}
  \end{center}
  \vspace{-0.55cm}
\end{figure}
The maximum transmission power at the $i$-th ST is given as $\beta_1=\beta_2=2$. The interference threshold at the $j$-th PR is set as $\gamma_1=\gamma_2=0.2$. The noise variance is $\sigma_{\!\scriptscriptstyle R}^2=\sigma_{\!\scriptscriptstyle E}^2=\sigma^2$.

Fig. 6 plots the average secrecy sum rate with BPSK, QPSK and 8PSK modulations. Results in Fig. 6 show that the average secrecy sum rate is an increasing function with respect to the order of modulation. They also indicate that our proposed precoding design can achieve robust performances for a large range of SNR with different modulations.

\section{Conclusion}
In this paper, we have considered the precoding design for the fading CMAC-WT with finite-alphabet inputs. We have presented a two-layer precoding algorithm, which exploits statistical CSI of fading channels, to maximize the approximated average secrecy sum rate. The key idea of our algorithm is to find a computationally efficient DC representation of the approximated average secrecy sum rate. By introducing a new matrix variable, we have reformulated the approximated average secrecy sum rate as a DC function, and then generated a sequence of relaxed sets to approach the non-convex feasible set. Each relaxed set can be expressed as the union of convex sets. Finally, near optimal precoding matrices have been obtained iteratively by maximizing the approximated average secrecy sum rate over a sequence of relaxed sets.

Several numerical results have been provided to demonstrate the efficacy of our proposed precoding algorithm. They have also shown that the proposed precoding algorithm is superior to the conventional Gaussian precoding method and no precoding case in the medium and high SNR regimes.

\appendices
\section*{Appendix A\\ Proofs of Propositions \textrm{1}-\textrm{2}}
\begin{IEEEproof}[Proof of Proposition 1]
We rewrite $\mathcal{F}_{\mathrm{init}}$ as the union of $K$ subsets
\begin{align}
\mathcal{F}_{\mathrm{init}}=\bigcup_{i=1}^{K}\mathcal{\bar{F}}_{i}
\end{align}
where $\mathcal{\bar{F}}_{i}$ is given by
\begin{align}
\mathcal{\bar{F}}_{i}=\big\{(\bQ,\bp)|(\bQ,\bp) \in \mathcal{F}_{\mathrm{init}}, \bp \in \mathcal{B}_i\big\}.
\end{align}
For any $(\bQ,\bp)\in \mathcal{\bar{F}}_{i}$, the following inequalities hold
\begin{align}
&(\bp-\mathbf{l}(\mathcal{B}_{i}))\cdot(\bp-\mathbf{l}(\mathcal{B}_{i}))^T \geq \bm{0}\\
&(\bp-\mathbf{u}(\mathcal{B}_{i}))\cdot(\bp-\mathbf{u}(\mathcal{B}_{i}))^T \geq \bm{0}\\
&(\bp-\mathbf{l}(\mathcal{B}_{i}))\cdot(\bp-\mathbf{u}(\mathcal{B}_{i}))^T \leq \bm{0}\\
&\bQ=\bp\bp^{T}, \mathbf{l}(\mathcal{B}_i)\leq\bp\leq\mathbf{u}(\mathcal{B}_i).
\end{align}
Thus, $\mathcal{\bar{F}}_{i}$ can be rewritten as
\begin{align}
\mathcal{\bar{F}}_{i}=\big\{(\bQ,\bp)|(\bQ,\bp) \in \mathcal{F}_{\mathrm{init}}, \bp \in \mathcal{B}_i\big\}\cap \mathcal{S}(\mathcal{B}_{i}).
\end{align}
By relaxing $\bQ=\bp\bp^T$ in $\mathcal{\bar{F}}_{i}$ into $\bQ\succeq\bp\bp^T$, one can easily obtain the following
\begin{align}
\mathcal{\bar{F}}_{i}\subseteq \mathcal{C}(\mathcal{B}_i), \; \forall i.
\end{align}
Therefore, $\mathcal{F}_{\mathrm{init}} \subseteq\mathcal{C}(\mathcal{B}_1) \cup ... \cup \mathcal{C}(\mathcal{B}_K)$. This completes the proof.
\end{IEEEproof}

\begin{IEEEproof}[Proof of Proposition 2]
We divide $\mathcal{S}(\mathcal{B})$ into two subsets
\begin{equation}
\begin{aligned}
\mathcal{S}(\mathcal{B})=\mathcal{S}_{1}(\mathcal{B}) \cup \mathcal{S}_{2}(\mathcal{B})
\end{aligned}
\end{equation}
where $\mathcal{S}_{1}(\mathcal{B})$ and $\mathcal{S}_{2}(\mathcal{B})$ are given by
\begin{align}
&\mathcal{S}_{1}(\mathcal{B})=\big\{(\bQ,\bp)|(\bQ,\bp) \in \mathcal{S}(\mathcal{B}), \bp \in \mathcal{B}_1\big\}\nonumber\\
&\mathcal{S}_{2}(\mathcal{B})=\big\{(\bQ,\bp)|(\bQ,\bp) \in \mathcal{S}(\mathcal{B}), \bp \in \mathcal{B}_2\big\}.
\end{align}
It is obvious that if we can prove
\begin{align}
&\mathcal{S}(\mathcal{B}_1) \subseteq \mathcal{S}_{1}(\mathcal{B})\nonumber\\
&\mathcal{S}(\mathcal{B}_2) \subseteq \mathcal{S}_{2}(\mathcal{B})
\end{align}
then $\mathcal{C}(\mathcal{B}_1) \cup \mathcal{C}(\mathcal{B}_2) \subseteq \mathcal{C}(\mathcal{B})$. We will restrict our attention to show $\mathcal{S}(\mathcal{B}_1) \subseteq \mathcal{S}_{1}(\mathcal{B})$, and $\mathcal{S}(\mathcal{B}_2) \subseteq \mathcal{S}_{2}(\mathcal{B})$ can be proved in the same way.

Since $\mathcal{B}_1\subseteq \mathcal{B}$, we have
\begin{align}
&\mathbf{l}(\mathcal{B}) \leq \mathbf{l}(\mathcal{B}_1)\leq \bu(\mathcal{B}_1) \leq \bu(\mathcal{B}).
\end{align}
Therefore, the following inequalities hold for any $\mathbf{l}(\mathcal{B}_1)\leq\bp\leq \bu(\mathcal{B}_1)$:
\begin{align*}
&[\mathbf{l}(\mathcal{B}_1\!)\!-\!\mathbf{l}(\mathcal{B})][\bp\!-\!\mathbf{l}(\mathcal{B}_1\!)]^T\!+\![\bp\!-\!\mathbf{l}(\mathcal{B})][\mathbf{l}(\mathcal{B}_1\!)\!-\!\mathbf{l}(\mathcal{B})]^T\!\geq\! \bm{0}\nonumber\\
&[\mathbf{u}(\mathcal{B}_1\!)\!-\!\mathbf{u}(\mathcal{B})][\bp\!-\!\mathbf{u}(\mathcal{B}_1\!)]^{T}\!\!+\![\bp\!-\!\mathbf{u}(\mathcal{B})][\mathbf{u}(\mathcal{B}_1\!)\!-\!\mathbf{u}(\mathcal{B})]^T\!\geq\! \bm{0}\nonumber\\
&[\mathbf{l}(\mathcal{B}_1\!)\!-\!\mathbf{l}(\mathcal{B})][\bp\!-\!\mathbf{u}(\mathcal{B})]^T\!\!+\![\bp\!-\!\mathbf{l}(\mathcal{B}_1\!)][\mathbf{u}(\mathcal{B}_1\!)\!-\!\mathbf{u}(\mathcal{B})]^T\!\leq\! \bm{0}.
\end{align*}
The above inequalities can be rewritten respectively as
\begin{align}\label{LSTI}
\quad\bQ\!-\!\bL_{\bp}(&\mathcal{B})\!-\!\bL_{\bp}(\mathcal{B})^T\!+\!\mathbf{l}(\mathcal{B})\cdot\mathbf{l}(\mathcal{B})^T\geq\nonumber\\
&\bQ\!-\!\bL_{\bp}(\mathcal{B}_1)\!-\!\bL_{\bp}(\mathcal{B}_1)^T\!+\!\mathbf{l}(\mathcal{B}_1)\cdot\mathbf{l}(\mathcal{B}_1)^T\nonumber\\
\quad\bQ\!-\!\bU_{\bp}(&\mathcal{B})\!-\!\bU_{\bp}(\mathcal{B})^T\!+\!\mathbf{u}(\mathcal{B})\cdot\mathbf{u}(\mathcal{B})^T\geq\nonumber\\
&\bQ\!-\!\bU_{\bp}(\mathcal{B}_1)\!-\!\bU_{\bp}(\mathcal{B}_1)^T\!+\!\mathbf{u}(\mathcal{B}_1)\cdot\mathbf{u}(\mathcal{B}_1)^T\nonumber\\
\bQ\!-\!\bL_{\bp}(&\mathcal{B})\!-\!\bU_{\bp}(\mathcal{B})^T\!+\!\mathbf{l}(\mathcal{B})\cdot\mathbf{u}(\mathcal{B})^T\leq\nonumber\\
&\bQ\!-\!\bL_{\bp}(\mathcal{B}_1)\!-\!\bU_{\bp}(\mathcal{B}_1)^T\!+\!\mathbf{l}(\mathcal{B}_1)\cdot\mathbf{u}(\mathcal{B}_1)^T
\end{align}
where $\bL_{\bp}(\mathcal{B})=\mathbf{l}(\mathcal{B})\cdot\bp^T$, and $\bU_{\bp}(\mathcal{B})=\mathbf{u}(\mathcal{B})\cdot\bp^T$. Inequalities \eqref{LSTI} provide a sufficient condition for $\mathcal{S}(\mathcal{B}_1)\subseteq\mathcal{S}_{1}(\mathcal{B})$. Therefore, $\mathcal{C}(\mathcal{B}_1) \cup \mathcal{C}(\mathcal{B}_2) \subseteq \mathcal{C}(\mathcal{B})$. This completes the proof.
\end{IEEEproof}

\section*{Appendix B\\ Proof of Proposition 3}
\begin{IEEEproof}
Since $\{\varphi(\mathcal{F}_k)\}$ is a monotonically decreasing sequence lower bounded by $\varphi(\mathcal{F}_{\mathrm{init}})$, the limit of $\{\varphi(\mathcal{F}_k)\}$ exists \cite{rudin1964principles}. Suppose that
\begin{align}
\lim_{k\rightarrow\infty} \varphi(\mathcal{F}_k)=v>\varphi(\mathcal{F}_{\mathrm{init}})
\end{align}
then for any $\varepsilon>0$, there exists $K>0$ such that for any $k>K$,
\begin{align} \label{conver1}
v<\varphi(\mathcal{C}(\mathcal{B}_g))<v+\varepsilon.
\end{align}
Let $r(\mathcal{B})$ denote the length of the longest edge of a hyperrectangle $\mathcal{B}$ satisfying $\mathcal{B}\subseteq \mathcal{B}_{\mathrm{init}}$. In each iteration of Algorithm 1, we divide $\mathcal{B}_g$ along $r(\mathcal{B}_g)$ into two hyperrectangles. Therefore, $r(\mathcal{B}_g)$ should satisfy the following condition:
\begin{align}\label{limitt}
\lim_{k \rightarrow \infty} r(\mathcal{B}_g)=0.
\end{align}
We further denote the center of $\mathcal{B}_g$ by $\bp_{g}$, i.e., $\bp_{g}=(\mathbf{l}(\mathcal{B}_g)+\bu(\mathcal{B}_g))/2$. When $r(\mathcal{B}_g)\rightarrow 0$, we have
\begin{align}
\mathcal{S}(\mathcal{B}_g)\rightarrow \{(\bQ,\bp)|\bQ=\bp_{g}\bp_{g}^{T}, \bp=\bp_{g}\}.
\end{align}
Therefore, $\mathcal{C}(\mathcal{B}_g)$ converges to a point when $\bp_{g}$ belongs to the feasible set $\mathcal{P}$, otherwise $\mathcal{C}(\mathcal{B}_g)$ is an empty set. Thus we have
\begin{align}\label{limittt}
\lim_{r(\mathcal{B}_g)\rightarrow 0}\varphi(\mathcal{C}(\mathcal{B}_g))=f(\bp_{g})-g(\bp_{g}), \quad \bp_g\in \mathcal{P}.
\end{align}
Combining \eqref{limitt} and \eqref{limittt}, we conclude that
\begin{align}
\lim_{k\rightarrow \infty}\varphi(\mathcal{C}(\mathcal{B}_g))=f(\bp_{g})-g(\bp_{g})<v
\end{align}
which is contradictory to \eqref{conver1}. Therefore, $\{\varphi(\mathcal{F}_k)\}$ converges to $\varphi(\mathcal{F}_{\mathrm{init}})$. This completes the proof.
\end{IEEEproof}
\maketitle
\bibliographystyle{IEEEtran}
\bibliography{IEEEabrv,reference}

\end{document}